\documentclass[onecollarge,natbib]{svjour2}
\bibpunct{[}{]}{;}{n}{}{,} % to get "[numbered]" references from natbib
\smartqed  % flush right qed marks, e.g. at end of proof
\usepackage{graphicx}
\usepackage{bm}
\journalname{Few-Body Systems}
\begin{document}

\title{Transverse densities of octet baryons from chiral effective field theory}


\author{Jose Manuel Alarc\'on   \and
        Astrid N.~Hiller Blin \and
        Christian Weiss}


\institute{Jose Manuel Alarc\'on and Christian Weiss \at
              Theory Center, Jefferson Lab, Newport News, VA 23606, USA \\
%              Tel.: +1-757-269-5385\\
%              Fax: +123-45-678910\\
              \email{alarcon@jlab.org}, \email{weiss@jlab.org}           %  \\
           \and
           Astrid N.~Hiller Blin \at
Departamento de F\'isica Te\'orica and IFIC, Centro Mixto Universidad de Valencia-CSIC, 
Institutos de Investigaci\'on de Paterna, E-46071 Valencia, Spain\\
%Tel.: +34-963-543528\\
\email{astrid.hiller@uv.es}
}

\date{%Received: date / Accepted: date
}

\maketitle

\begin{abstract}
Transverse densities describe the distribution of charge and current at fixed light-front 
time and provide a frame-independent spatial representation of hadrons as relativistic systems. 
We calculate the transverse densities of the octet baryons at peripheral distances $b = O(M_\pi^{-1})$ 
in an approach combining chiral effective field theory ($\chi$EFT) and dispersion analysis.
The densities are represented as dispersive integrals of the imaginary parts of the
baryon electromagnetic form factors in the timelike region (spectral functions).
The spectral functions on the two-pion cut at $t > 4 M_\pi^2$ are computed using relativistic $\chi$EFT 
with octet and decuplet baryons in the EOMS renormalization scheme. The calculations are extended into 
the $\rho$-meson mass region, using a dispersive method that incorporates the timelike pion form-factor data. 
The approach allows us to construct densities at distances $b > 1$ fm with controlled uncertainties. 
Our results provide insight into the peripheral structure of nucleons and hyperons 
and can be compared with empirical densities and lattice-QCD calculations.
\end{abstract}

\section{Introduction}
\label{intro}
Light-front quantization offers a natural framework for formulating the 
spatial structure of relativistic systems and exploring its connection
to the underlying dynamics; see \cite{Brodsky:1997de} for a review. 
In this framework, hadronic current matrix
elements (vector, axial) are represented in terms of transverse densities,
which are two-dimensional Fourier transforms of the invariant form factors 
and describe the transverse spatial distribution of charge and current 
in the hadron at fixed light-front time $x^+ \equiv x^0 + x^3$ 
\cite{Soper:1976jc,Burkardt:2000za,Miller:2007uy,Miller:2010nz}. 
The transverse densities are frame independent (they are invariant under longitudinal
boosts and transform kinematically under transverse boosts) and thus provide 
an objective spatial representation of the hadron as a relativistic system.
In composite models of hadron structure they correspond to proper densities of 
the light-front wave functions of the system. In the context of QCD the transverse 
charge and current densities in hadrons can be related to the generalized parton 
distributions (GPDs) describing the distribution of quarks and antiquarks in 
longitudinal momentum and transverse position \cite{Burkardt:2000za,Burkardt:2002hr}.
The charge and magnetization densities in the nucleon have been extracted from the
available electromagnetic form-factor data \cite{Perdrisat:2006hj,Punjabi:2015bba} 
and provide interesting insight into the nucleon structure; see \cite{Miller:2010nz} 
for a review. It is worthwhile to explore how the transverse densities could be 
calculated using theoretical methods, and how the studies could be extended to 
other baryons.

The transverse densities at a given distance can be connected with the ``exchange mechanisms'' 
acting in the hadron form factors --- virtual processes in which the current couples to 
the hadron through the exchange of a hadronic system in the $t$-channel. This connection
can be made rigorous in a dispersive representation of the transverse 
densities \cite{Strikman:2010pu,Granados:2013moa}.
At distances $b = O(M_\pi^{-1})$, the densities are governed by soft-pion exchange
between the current and the hadron and can be calculated model-independently using
chiral effective field theory ($\chi$EFT). 
Detailed studies of the peripheral densities in the nucleon were performed
in \cite{Granados:2013moa,Granados:2015rra,Granados:2015lxa} using $\chi$EFT with 
SU(2) flavor group, and a simple quantum-mechanical
interpretation of the results was obtained (chiral light-front wave functions, orbital 
motion of a peripheral pion). At distances $b \sim 1$ fm, the transverse densities are 
dominated by vector-meson exchange (isovector $\rho$; isoscalar $\omega, \phi$) and
offer interesting insight into the duality between quark structure and meson 
exchange \cite{Miller:2011du}. 

In these proceedings we report about a study of the peripheral densities of the
SU(3) flavor-octet baryons combining methods of $\chi$EFT and dispersion analysis.
The densities are represented as a dispersive integral over the imaginary parts of the
baryon electromagnetic form factors in the timelike region (spectral functions).
The isovector spectral functions on the two-pion cut at $t > 4 M_\pi^2$ are computed using 
relativistic $\chi$EFT with octet and decuplet baryons in the EOMS renormalization scheme. 
The calculations are extended into the $\rho$-meson mass region using a dispersive method 
that incorporates the timelike pion form-factor data. The methods allow us to construct the
densities down to distances $b \sim 1$ fm with controlled uncertainties. Details will 
be reported in a forthcoming publication \cite{inprep}.
\section{Transverse densities and dispersive representation}
\label{Stheoff}
The matrix element of the electromagnetic current between states of a flavor-octet baryon 
($B = N, \Sigma, \Xi$) with 4-momenta $p$ and $p'$ is described by two invariant form factors, 
$F_1^B(t)$ and $F_2^B(t)$ (the Dirac and Pauli form factors; we follow the conventions 
of \cite{Granados:2013moa}).
They are functions of the invariant momentum transfer $t = \Delta^2 = (p' - p)^2$, and can be 
measured and interpreted without specifying the form of relativistic dynamics or the reference frame. 
In the context of light-front quantization, one considers the current matrix element 
in a frame where the 4-momentum transfer has only transverse components 
$\Delta^+ = \Delta^- = 0, \vec{\Delta}_T \neq 0$, and represents the form factors as Fourier transforms
of certain two-dimensional densities (here $b \equiv |\vec{b}|$),
\begin{equation}
\label{Eq:rho_def}
F_i^B(t = -|\vec{\Delta}_T|^2) \;\; = \;\; \int \mathrm{d}^2 b \; 
e^{i \vec{\Delta}_T \cdot \vec{b}} \; \rho_i^B (b) 
\hspace{2em} (i=1,2).
\end{equation}
The interpretation of $\rho_{1, 2} (b)$ as spatial densities is discussed 
in \cite{Burkardt:2000za,Miller:2010nz} and summarized in \cite{Granados:2013moa}.
Their spatial integral reproduces the total charge and anomalous magnetic moment of the baryon.
In a state where the baryon is localized in transverse space at the origin, $\rho_1^B(b)$ describes 
the spin-independent $J^+$ current at light-front time $x^+ = 0$ and transverse position $\vec{b}$, while 
$\widetilde\rho_2^B (b) \equiv \partial / \partial b [\rho_2^B(b) /(2 M_B)]$ describes the spin-dependent 
part of the current in a transversely polarized nucleon. 

The form factors are analytic functions of $t$ and satisfy unsubtracted dispersion relations
\begin{equation}
F_i^{B} (t) \;\; = \;\; 
\int_{4M_\pi^2}^\infty \frac{\mathrm{d}t'}{t' - t - i0} 
\; \frac{\textrm{Im}\, F_i^{B} (t')}{\pi}
\hspace{2em} (i=1,2).
\label{Eq:ff-spectral-rep}
\end{equation}
The spectral functions are given by the imaginary parts of the form factors on the principal cut
starting at $t = 4 M_\pi^2$. They describe processes in which the current at timelike $t > 0$
converts to a hadronic state that couples to the $B\bar B$ system. Most of the relevant processes
are in the unphysical region below the two-baryon threshold at $t = 4 M_B^2$, where the spectral 
function can only be computed theoretically. From Eqs.~(\ref{Eq:rho_def}) and (\ref{Eq:ff-spectral-rep}),
one obtains a dispersive representation of the transverse densities \cite{Strikman:2010pu}
\begin{eqnarray}
\rho_1^B(b) &=& \phantom{-} \int_{4M_\pi^2}^{\infty}\hspace{-0.4cm} \mathrm{d}t\ 
\frac{K_0(\sqrt{t}b)}{ 2\pi}  \frac{\mathrm{Im}F_1^B(t)}{\pi},
\label{disp_rho_1}
\\[0ex]
\widetilde{\rho}_2^B(b) &=& - \int_{4M_\pi^2}^{\infty}\hspace{-0.4cm} \mathrm{d}t\ \frac{\sqrt{t} 
K_1(\sqrt{t}b)}{ 4\pi M_B} \frac{\mathrm{Im}F_2^B(t)}{\pi},
\label{disp_rho_2}
\end{eqnarray}
where $K_n (n = 0, 1)$ are the modified Bessel functions. The integrals in Eqs.~(\ref{disp_rho_1}) 
and (\ref{disp_rho_2}) converge exponentially at large $t$, $K_n(\sqrt{t} b) \sim \exp(-\sqrt{t}b)$ 
at $\sqrt t b \gg 1$, strongly suppressing contributions from high-mass hadronic states. Depending
on the distance $b$, the integrals sample the spectral functions in different regions of 
$t$ \cite{Miller:2011du}.
At $b > 2$ fm, the integrals extend over the near-threshold region $t = 4 M_\pi^2 + \textrm{few} \, M_\pi^2$, 
where the spectral functions arise from soft two-pion exchange between the baryon and the current. 
At $b \sim 1$ fm, the dominant contributions come from the vector-meson region of the spectral functions
$t \sim M_V^2$ ($V = \rho^0, \omega, \phi$). At even shorter distances, the integrals extend over
the high-mass region $t > 1$ GeV$^2$, where the spectral functions involve multi-hadron states and
are poorly known at present \cite{Belushkin:2006qa}. 
The dispersive representation Eqs.~(\ref{disp_rho_1}) and (\ref{disp_rho_2}) 
thus establishes a quantitative connection between the transverse densities and the exchange mechanisms 
in the form factor.
\section{Spectral functions from chiral EFT and dispersion theory}
\begin{figure*}
\begin{center}
\includegraphics[width=0.18\textwidth]{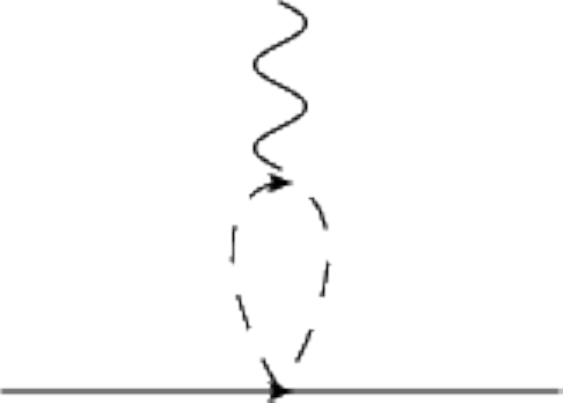}\label{Fig:LoopsFFi}
\hspace{5mm}
\includegraphics[width=0.18\textwidth]{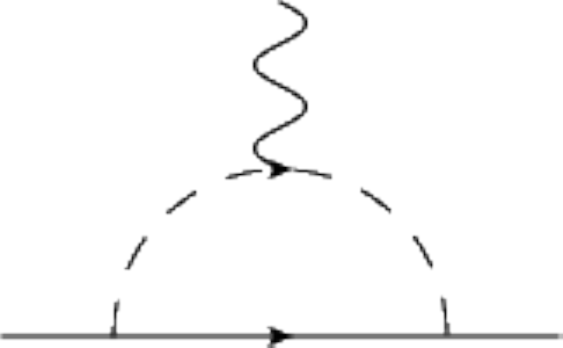}\label{Fig:LoopsFFa}
\hspace{5mm}
\includegraphics[width=0.18\textwidth]{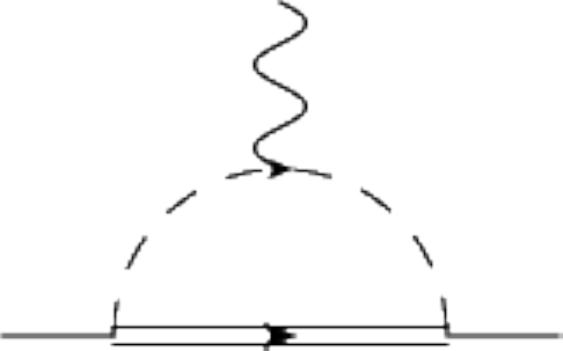}\label{Fig:LoopsFFc}
\end{center}
\caption{Chiral processes contributing to the spectral functions of the baryon form factors 
$\textrm{Im} F_{1, 2}^B(t)$ at $\mathcal{O}(p^3)$. Dashed lines: pions/kaons. 
Solid lines: octet baryons. Double lines: decuplet baryons.}
\label{Fig:LoopsOctet}
\end{figure*}
We compute the isovector spectral functions of the SU(3) octet-baryon form factors above the two-pion 
threshold using relativistic $\chi$EFT. Decuplet baryons are included as dynamical degrees of freedom 
within the small-scale expansion (SSE) \cite{Hemmert:1996xg,Hemmert:1997ye} (the first form factor 
calculations in this approach were performed in \cite{Bernard:1998gv}). The fields and Lagrangian 
are described in \cite{Geng:2009hh,Geng:2009ys,Ledwig:2014rfa}. The renormalization of the divergent 
pieces is performed within the EOMS scheme, which permits consistent power counting \cite{Fuchs:2003qc}.
The spectral functions arise from diagrams with a two-pion cut; the $O(p^3)$ diagrams are shown in
Fig.~\ref{Fig:LoopsOctet} (the imaginary parts are actually finite at this order and do not require 
renormalization). The low-energy constants at this order are given by the nucleon's axial charge
(or $\pi NN$ coupling) and the $\pi N \Delta$ coupling, so that the spectral functions are $\chi$EFT 
predictions free of unknown parameters. While not shown here, we have calculated the entire form factor from 
the full set of $\mathcal{O}(p^3)$ diagrams in order to verify gauge invariance, and reproduce the SU(2) 
results of \cite{Ledwig:2014rfa}.

\begin{figure*}[t]
\parbox[c]{.45\textwidth}{\includegraphics[width=.45\textwidth]{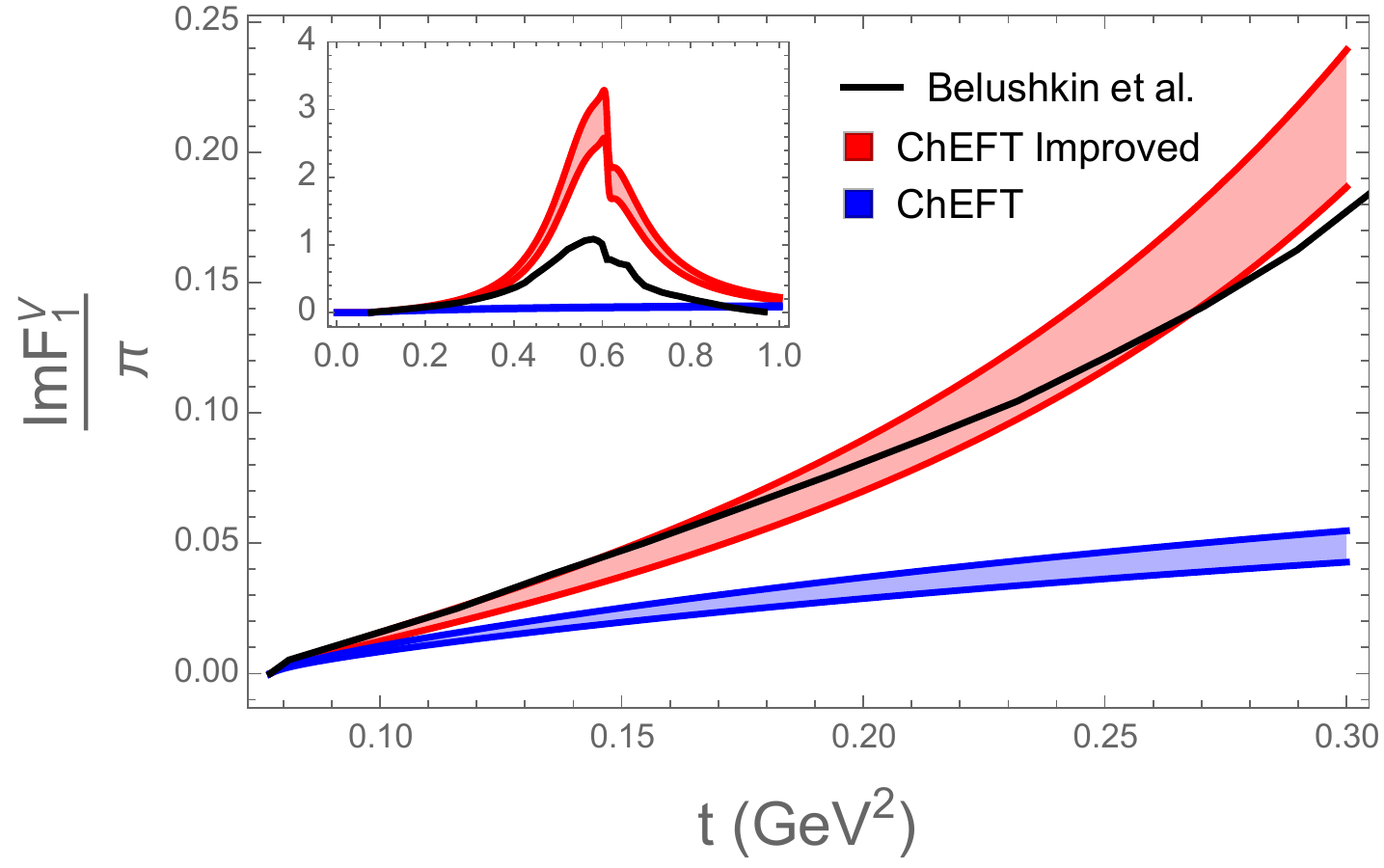}}
\hspace{.05\textwidth}
\parbox[c]{.4\textwidth}{
\caption{Dispersive improvement of the $\chi$EFT spectral function $\textrm{Im}\, F_1 (t)$,
Eq.~(\ref{Eq:improvement}). The main figure shows the spectral functions up to $t = 0.3$ GeV$^2$,
the inset figure shows them up to 1 GeV$^2$. Blue band: $\chi$EFT result at $O(p^3)$, including
$N$ and $\Delta$ intermediate states. Red band: Improved spectral function Eq.~(\ref{Eq:improvement}).
Black line: Dispersion-theoretical spectral function from \cite{Belushkin:2005ds}.
\label{fig:spectral}
}}
\end{figure*}

The $\chi$EFT expressions by themselves describe the baryon spectral functions only in the near-threshold 
region $t = 4 M_\pi^2 + \textrm{few} \, M_\pi^2$. In order to extend the description to higher $t$,
we use a dispersive technique following \cite{Frazer:1960zza,Frazer:1960zzb,Hohler:1974ht}; see
also \cite{Hammer:2003qv}. On general grounds, the isovector spectral function on the two-pion cut 
(neglecting the contributions from $4\pi$ states) can be expressed as
\begin{equation}
\label{Eq:Disp_rep_ImFV}
\textrm{Im} F_i^B(t) = \frac{k_{\rm cm}^3}{\sqrt{t}} \; \Gamma_i^B(t) \; F^*_{\pi}(t) 
\hspace{2em} (i=1,2),
\end{equation}
where $k_{\rm cm}=\sqrt{t/4 - M_\pi^2}$ is the $t$-channel center-of-mass momentum, $\Gamma_i^B(t)$ is 
the $I=J=1$  $\pi \pi \rightarrow \bar{B}B$ partial-wave amplitude, and $F_{\pi}(t)$ is the
pion form factor. The expression on the right-hand side of Eq.~(\ref{Eq:Disp_rep_ImFV}) is real 
because the complex functions $\Gamma_i^B(t)$ and $F_\pi(t)$ have the same phase on the two-pion 
cut (Watson theorem). It is convenient to rewrite Eq.~(\ref{Eq:Disp_rep_ImFV}) as
\begin{equation}
\label{Eq:Disp_rep_ImFV_ratio}
\textrm{Im} F_i^B(t) = \frac{k_{\rm cm}^3}{\sqrt{t}} \; \frac{\Gamma_i^B(t)}{F_{\pi}(t) } \; 
|F_\pi(t)|^2 \hspace{2em} (i=1,2).
\end{equation}
This representation has two major advantages: (i) The function $\Gamma_i^B(t)/F_{\pi}(t)$ has no
two-pion cut, because it is real at $t > 4 M_\pi^2$; (ii) the squared modulus 
$|F_\pi(t)|^2$ can be extracted directly from the 
$e^+e^- \rightarrow \pi^+\pi^-$ exclusive annihilation cross section, without 
determining the phase of the complex function. We now use $\chi$EFT to calculate the real function
$\Gamma_i^B(t)/F_{\pi}(t)$ at $t > 4 M_\pi^2$, and multiply with the empirical $|F_\pi(t)|^2$
containing the $\rho$ meson resonance. At $O(p^3)$ one has $F_\pi(t) \equiv 1$, so that the $\chi$EFT
result for the ratio $\Gamma_i^B(t)/F_{\pi}(t)$ is the same as that for the amplitude $\Gamma_i^B(t)$ itself,
and the prescription simply amounts to multiplying the $O(p^3)$ result for the spectral function by 
$|F_\pi(t)|^2$,
\begin{equation}
\label{Eq:improvement}
\textrm{Im} F_i^B(t) \, [\textrm{improved}] \;\; = \;\;
\textrm{Im} F_i^B(t) \, [O(p^3)] \; \times \; |F_\pi(t)|^2 \hspace{2em} (i=1,2).
\end{equation}
The prescription Eq.~(\ref{Eq:improvement}) results in a marked improvement of the $\chi$EFT 
predictions for the spectral functions. The improved $\chi$EFT results for the
nucleon ($B = N$) reproduce the dispersion-theoretical spectral functions 
(obtained by analytic continuation of the $\pi N$ phase shifts \cite{Hohler:1976ax,Belushkin:2005ds,Hoferichter:2016duk}) 
up to $t \sim 16 M_\pi^2$ within errors, and have
qualitatively the correct behavior even in the $\rho$-meson mass region (see Fig.~\ref{fig:spectral}).
We use this method to calculate the other octet-baryon spectral functions on the two-pion cut.
A detailed discussion of the procedure and its applications will be presented in \cite{inprep}.
\section{Peripheral transverse densities}
\label{Sresff}
Using the improved $\chi$EFT results for the spectral functions, we calculate the peripheral
isovector densities in the octet baryons. The restriction to distances $b > 1$ fm ensures that
the dispersion integrals Eqs.~(\ref{disp_rho_1}) and (\ref{disp_rho_2}) extend only over the
region of $t$ where our approximation to the spectral functions is justified. The uncertainties 
of the isovector densities are estimated by propagating the theoretical uncertainty of the 
spectral functions \cite{inprep}. 
In order to estimate also the isoscalar densities, we parametrize the isoscalar spectral functions 
of the octet baryons in the mass region $t < 1$ GeV$^2$ by vector-meson poles ($V = \omega, \phi$). 
The $VBB$ couplings are obtained from SU(3) symmetry, with certain assumptions regarding the $F/D$ ratio 
and the empirical $VNN$ couplings (we do not aim for a precise description of the isoscalar sector 
here, as the peripheral densities are dominated by the isovector component).

\begin{figure*}[t]
\begin{center}
\includegraphics[width=0.31\textwidth]{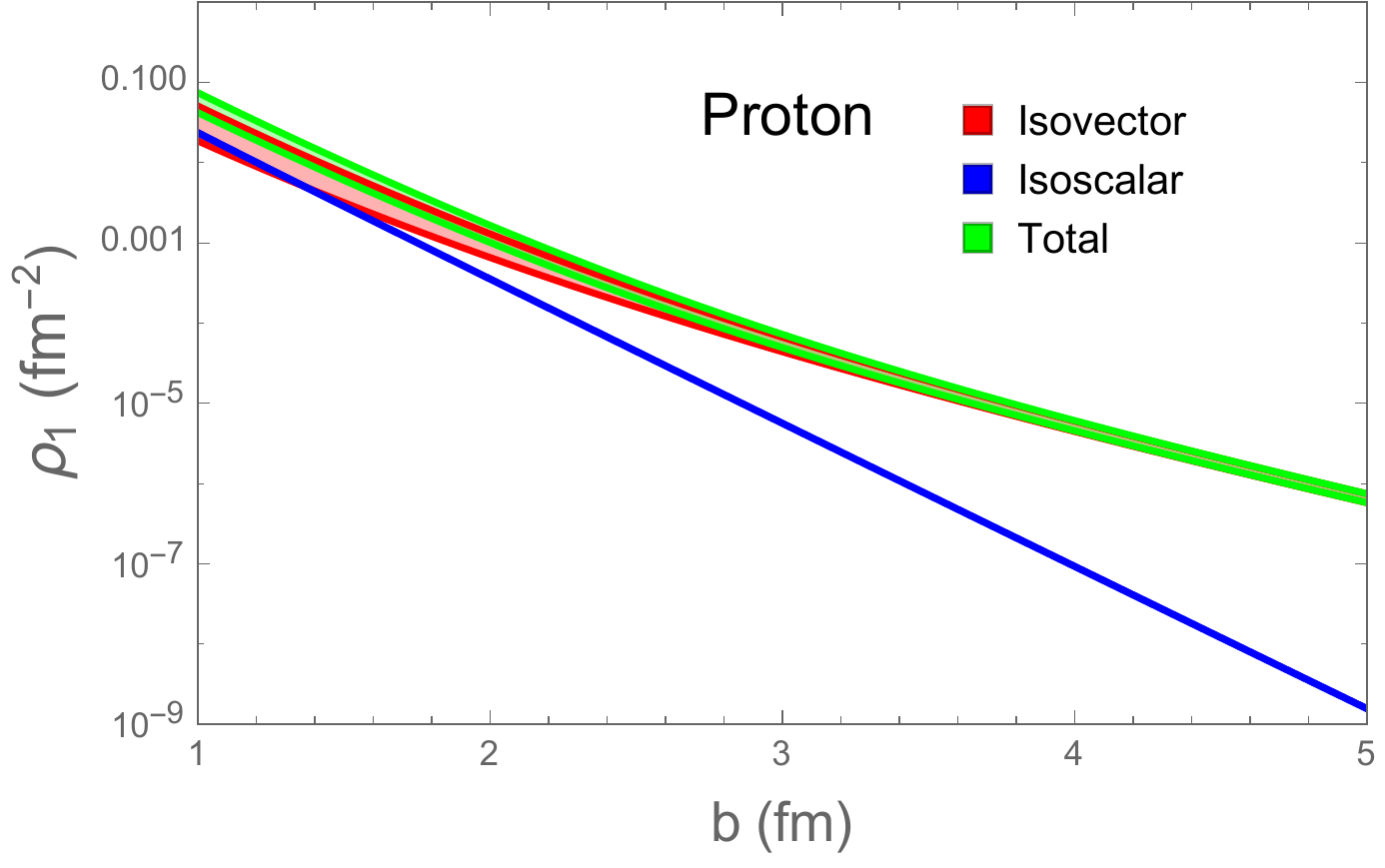}
\includegraphics[width=0.31\textwidth]{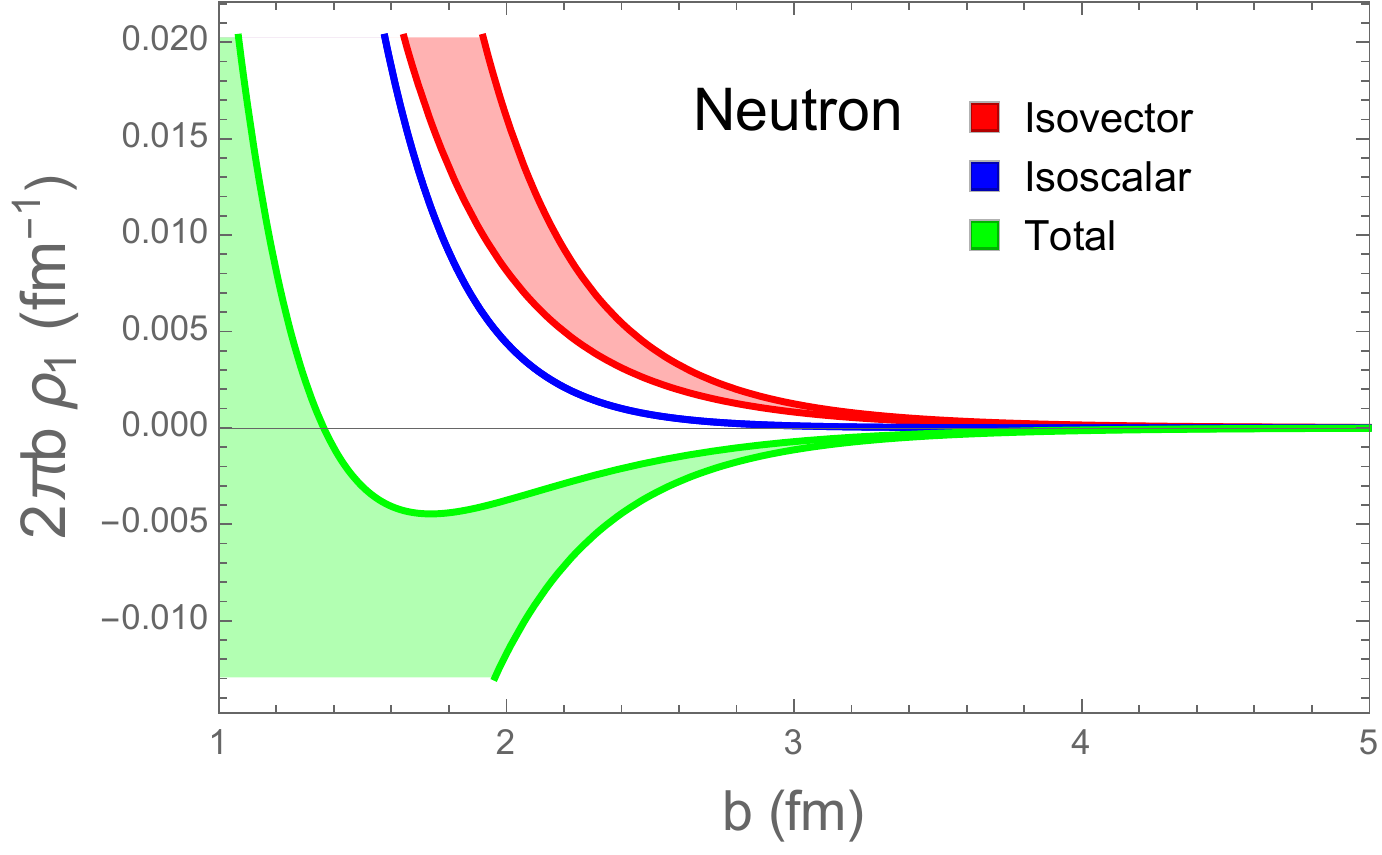}
\includegraphics[width=0.31\textwidth]{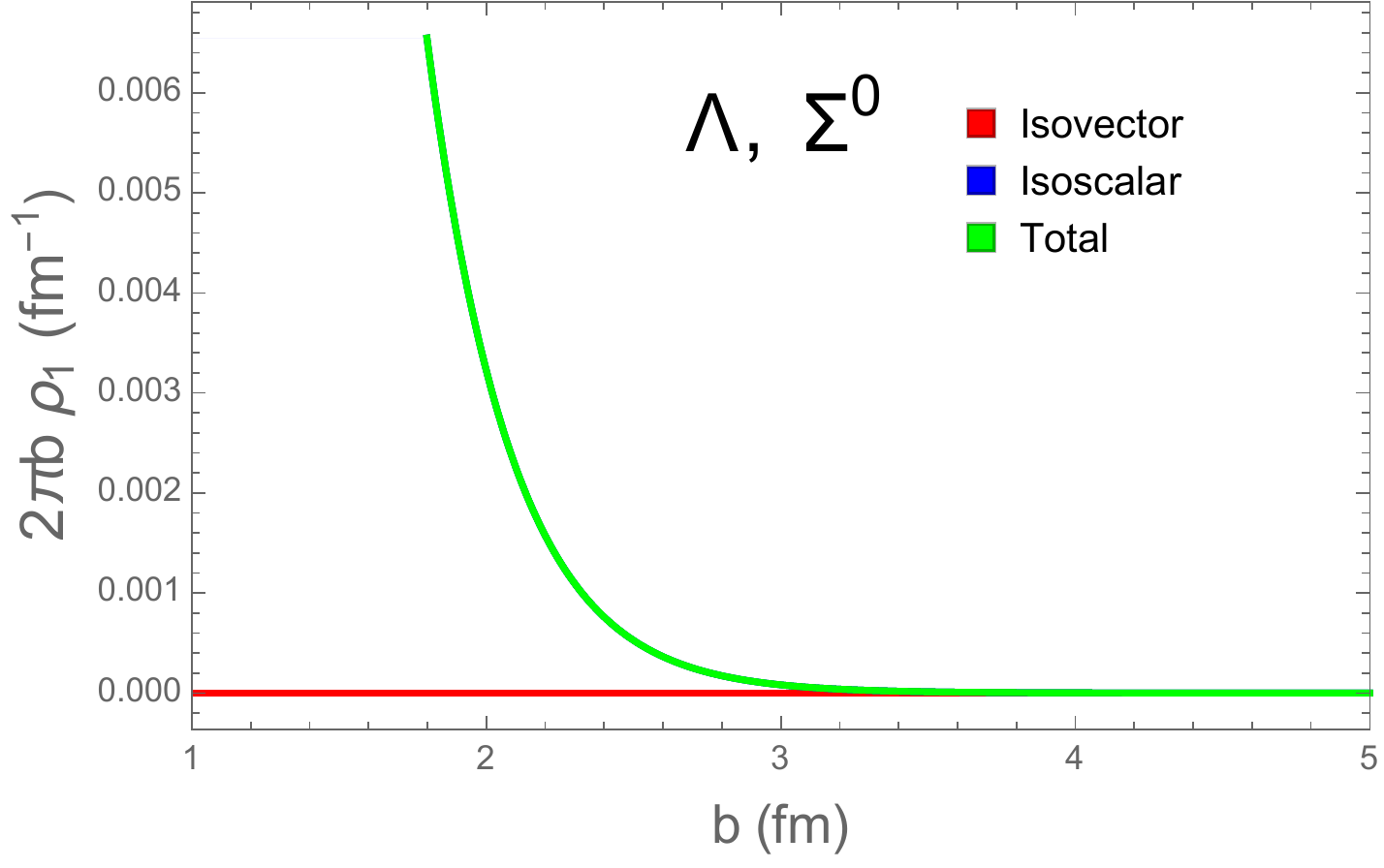} \\
\includegraphics[width=0.31\textwidth]{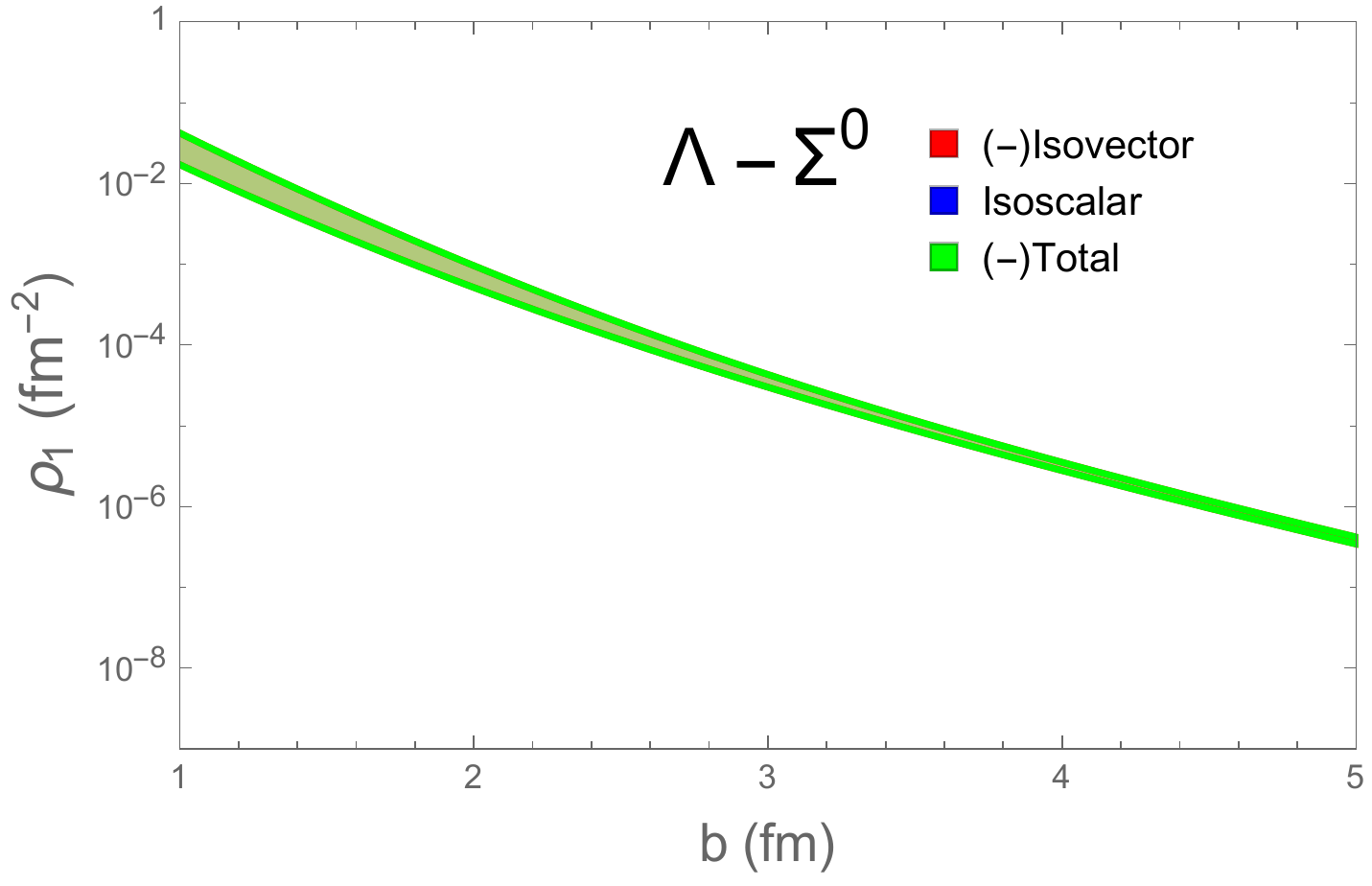}
\includegraphics[width=0.31\textwidth]{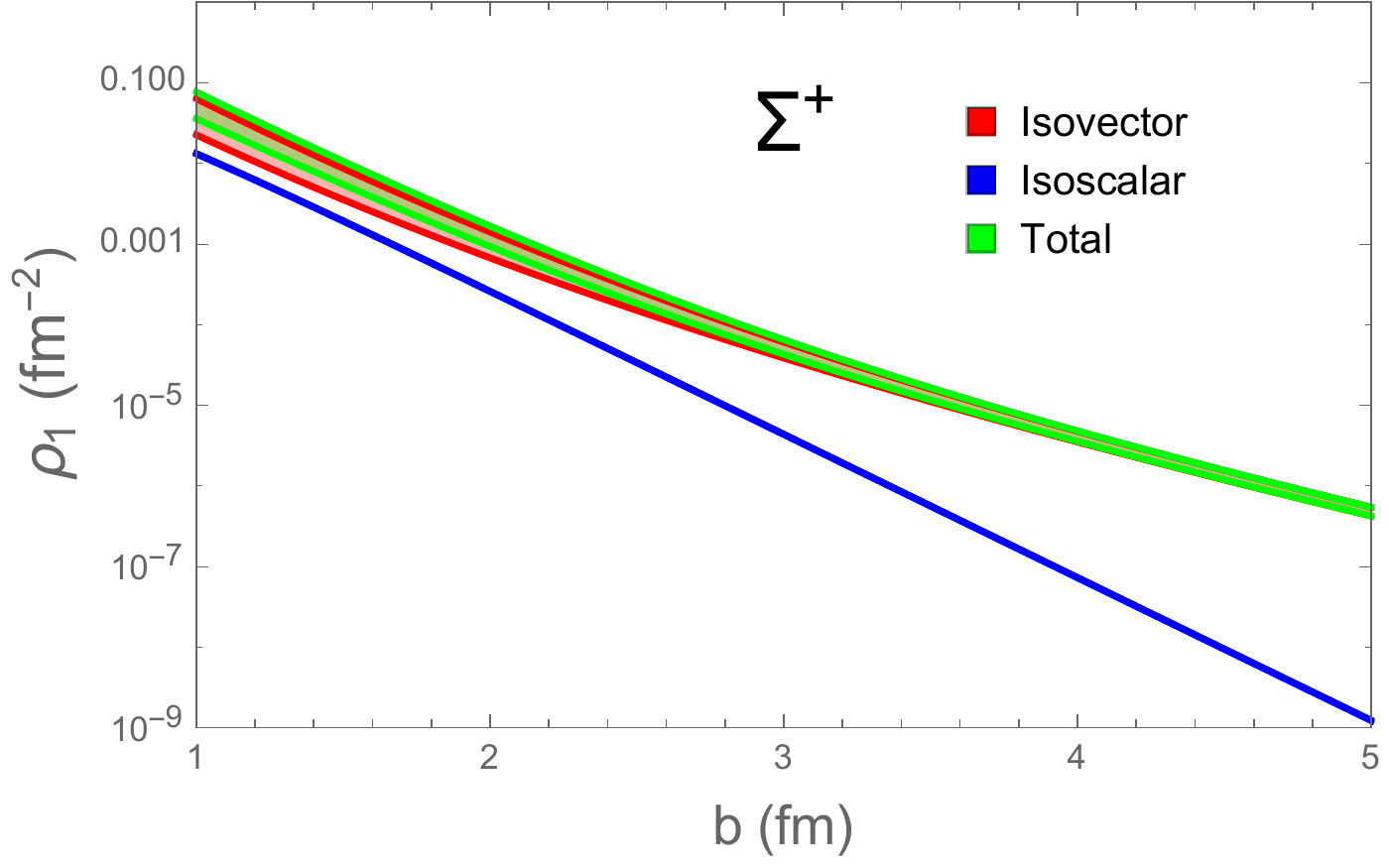}
\includegraphics[width=0.31\textwidth]{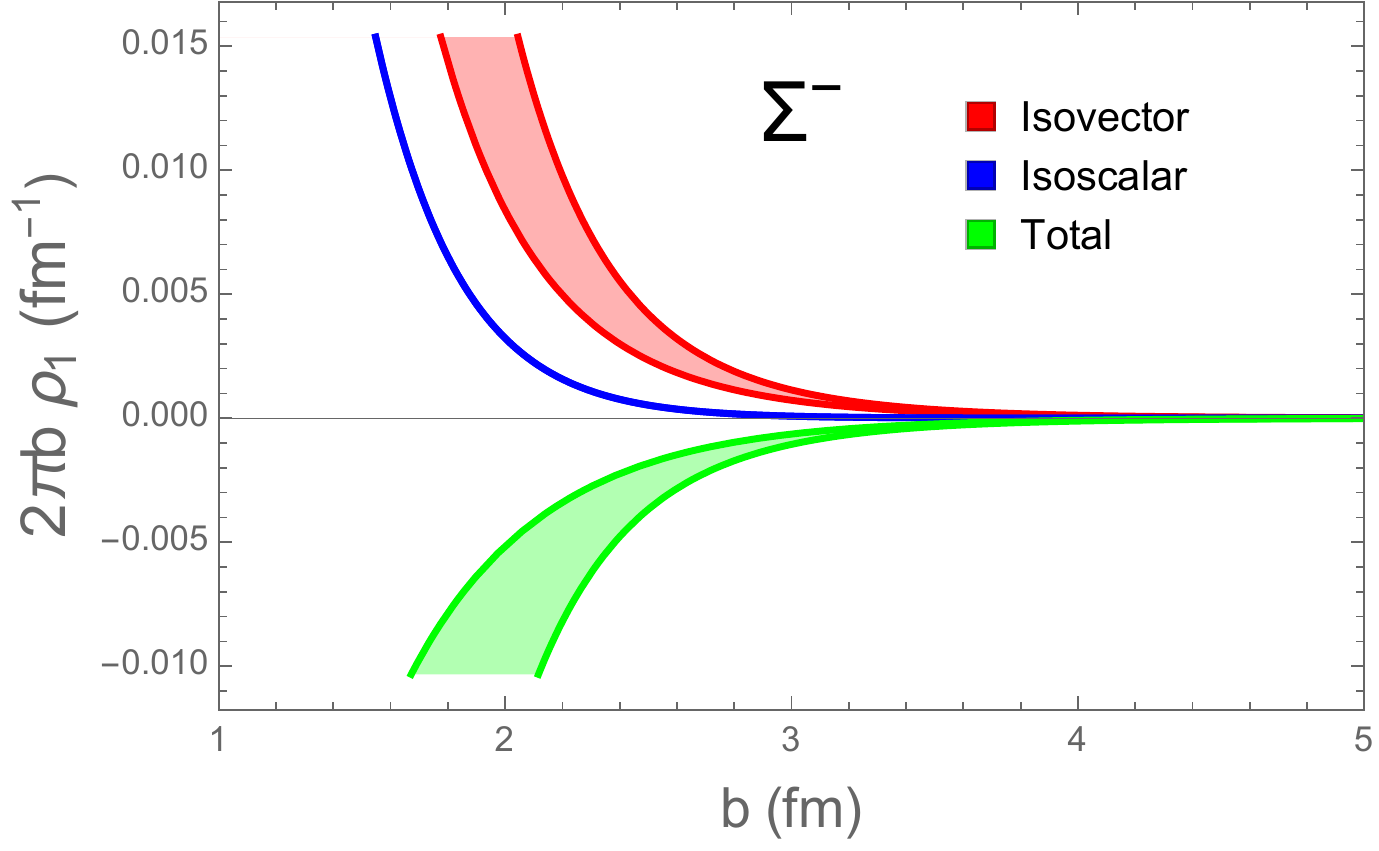} \\
\includegraphics[width=0.31\textwidth]{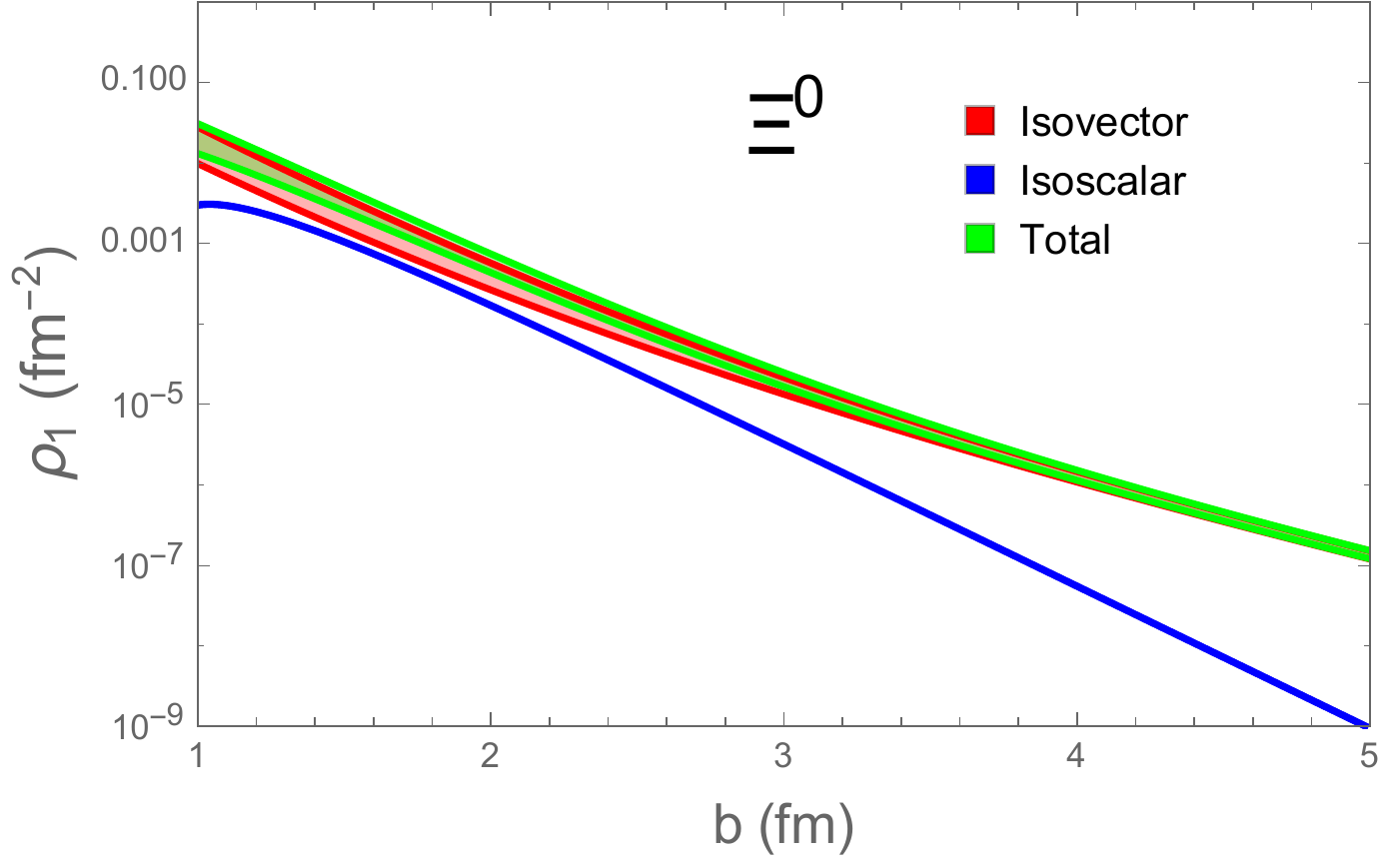}
\includegraphics[width=0.31\textwidth]{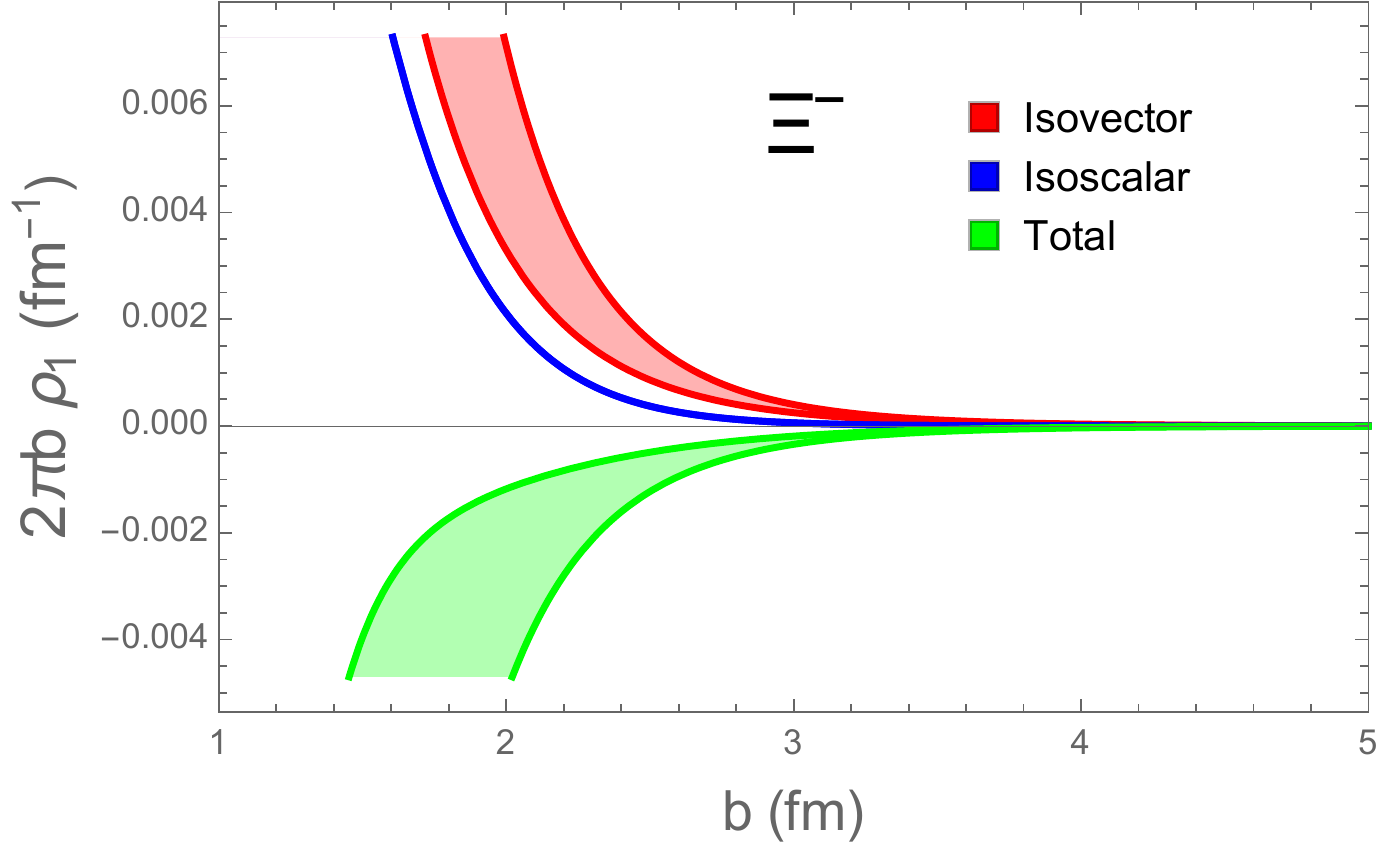}
\caption{Transverse charge densities of the octet baryons. Red: Isovector component
calculated using $\chi$EFT and dispersive improvement. Blue: Isoscalar component 
estimated from vector-meson poles. Green: Total density (sum or difference of isoscalar
and isovector components). For the densities with fixed sign (positive or negative) we
plot $\rho_1(b)$ on a logarithmic scale; for those with changing sign we plot the
radial densities $2\pi b \rho_1(b)$ on a linear scale.}
\label{fig:rho1_octet}
\end{center}
\end{figure*}

Results for the transverse charge densities are shown in Fig.~\ref{fig:rho1_octet}. The densities
decay exponentially at large $b$, as dictated by the analytic properties of
Eqs.~(\ref{disp_rho_1}) and (\ref{disp_rho_2}).
The comparison of the various components reveals several interesting features. At large distances 
$b > 3$ fm, the densities are dominated by the isovector component, resulting from two-pion exchange
near threshold. The isoscalar component generally becomes comparable to the isovector at distances 
below $b \sim 2$ fm, due to the similar strength of the spectral functions in the vector-meson 
region ($\rho$ vs.\ $\omega, \phi$). In the neutron, the isovector component dominates
for $b > 1.5$ fm, and causes the peripheral charge density to be negative. Studies of empirical neutron
densities have shown that at $b\sim 1$ fm the isoscalar takes over, and the neutron charge density becomes 
positive \cite{Miller:2007uy,Miller:2011du}; at such distances the present theoretical calculation 
has large uncertainties and cannot predict the sign.

In the $\Lambda$ and $\Sigma^0$ charge densities, the isovector component is absent because of 
the isospin selection rules for the $t$-channel process current $\rightarrow$ $B\bar B$:
the transitions $|I=1, I^3=0\rangle \rightarrow |I=0, I^3=0\rangle |I=0, I^3=0\rangle$
and $|I=1, I^3=0\rangle \rightarrow |I=1, I^3=0\rangle |I=1, I^3=0\rangle$ are both forbidden.
In contrast, in the $\Lambda\Sigma^0$ transition density, the isoscalar component is absent
and the density is pure isovector. The peripheral densities of the $\Lambda$ and $\Sigma^0$ 
thus provide a means to isolate the low-mass isovector and isoscalar exchanges in the form
factors. The $\Xi^0$ and $\Xi^-$ densities are similar to the proton and neutron in that 
both isovector and isoscalar exchanges are present. The isovector component is relatively 
smaller in the $\Xi$ states, due to the suppression of octet intermediate states in the pion 
loops for the $\Xi$ form factor (they are of the same order as the decuplet intermediate
states).

Similar features are exhibited by the magnetization densities of the octet baryons.
Using the same techniques we can also calculate the flavor decomposition of 
the peripheral densities and study the contribution of individual quark flavors to
the charge and magnetization \cite{inprep}.

\section{Summary and outlook}
\label{Ssummff}
We have presented a new approach to the calculation of the peripheral transverse densities using a combination
of relativistic $\chi$EFT and dispersion analysis. The dispersive improvement extends
the $\chi$EFT calculations of the isovector spectral functions into the vector-meson mass 
region and allows us to compute densities down to distances $\sim 1$ fm with controlled accuracy. 
The approach can be extended to the octet-baryon form factors of other operators (energy-momentum 
tensor, moments of GPDs). It can also be used to calculate the
transverse densities of the decuplet baryons (especially the $\Delta$ isobar), which are
studied in Lattice QCD, as well as to the densities of the octet-decuplet transition
form factors (especially $N \rightarrow \Delta$), which are measured in resonance electroproduction.

Our approach consistently includes the contributions from decuplet intermediate states in the 
$\pi/K$ loop diagrams. These contributions are numerically important in the transverse densities at
distances $b \sim$ 1--2 fm. They are also essential for ensuring the proper scaling behavior
of the densities in the large--$N_c$ limit of QCD. This was demonstrated for the SU(2) $\chi$EFT
in \cite{Granados:2013moa,Granados:2016jjl}, and can be shown for the present SU(3) 
calculation as well.

\begin{acknowledgements}
This material is based upon work supported by the U.S.~Department of Energy, Office of Science, 
Office of Nuclear Physics under contract DE-AC05-06OR23177.
This work was supported by the Spanish Ministerio de Econom\'{\i}a y Competitividad (MINECO)
and the European fund for regional development (FEDER) under contracts FIS2014-51948-C2-2-P and SEV-2014-0398.
\end{acknowledgements}

% BibTeX users please use
%\bibliographystyle{spbasic}
%\bibliography{mybib}   % name your BibTeX data base

\end{document}